\documentclass[12pt,letterpaper]{article}

\usepackage{graphicx,array}
\usepackage{url}
\usepackage{color}
\usepackage{latexsym}
\usepackage{amsthm}
\usepackage{amsmath}
\usepackage{amssymb}
\usepackage{amsfonts}
\usepackage[numbers,sort&compress]{natbib}
\usepackage{bm}
\usepackage{slashed}
\usepackage{mathrsfs}
\usepackage{enumerate}
\usepackage{tikz}
\usepackage{siunitx}
\usepackage{mdframed}
\usepackage{setspace}  
\usepackage{esvect}
\usepackage{esint}
\usepackage{subcaption}
\usepackage[normalem]{ulem}
\usepackage{setspace}
\usepackage{multirow}
\usepackage{diagbox}
\usepackage{soul}
\usepackage{multirow}

\usepackage[version=4]{mhchem}

\usepackage[utf8x]{inputenc}%
\usepackage{tcolorbox}%
\usepackage[normalem]{ulem} 

\usepackage{hyperref} 
\hypersetup{
    colorlinks=true,       
    linkcolor=red,          
    citecolor=blue,        
    filecolor=magenta,      
    urlcolor=blue           
}
\usepackage[all]{hypcap} 

\usepackage{natbib}
\newcommand{\nc}{\newcommand}

\nc{\beq}{\begin{equation}}  
\nc{\eeq}{\end{equation}}  
\nc{\beqa}{\begin{eqnarray}}  
\nc{\eeqa}{\end{eqnarray}}  
\nc{\bit}{\begin{itemize}}  
\nc{\eit}{\end{itemize}}  

\newcommand{\eg}{{\it e.g.}}
\newcommand{\ie}{{\it i.e.}}

\usepackage{floatrow}
\newfloatcommand{capbtabbox}{table}[][\FBwidth]

\usepackage{blindtext}

\def\figureautorefname~#1\null{Fig.\,#1\null}
\def\tableautorefname~#1\null{Tab.\,#1\null}

\def\equationautorefname~#1\null{Eq.\,(#1)\null}

\def\lambdaX{\lambda_{\rm X}}
\def\EX{E_{\rm X}}

\title{
{\bf IceCube at the Frontier of Macroscopic Dark Matter Direct Detection }
	\author{\large Yang Bai$^{\,\star}$, Joshua Berger$\,^\diamond$, and Mrunal Korwar$^{\,\star}$}
	\date{\small \it 
		$^\star$Department of Physics, University of Wisconsin-Madison, Madison, WI 53706, USA\\
		$^\diamond$Colorado State University, Fort Collins, CO 80523, USA \\
	    }
}

\begin{document}

\maketitle

\setlength{\parskip}{0.2ex}

\begin{abstract}	
For a class of macroscopic dark matter models, inelastic scattering of dark matter off a nucleus can generate electromagnetic signatures with GeV-scale energy. The IceCube detector, with its kilometer-scale size, is ideal for directly detecting such inelastic scattering. Based on the slow particle trigger for the DeepCore detector, we perform a detailed signal and background simulation to estimate the discovery potential. For order 1 GeV deposited energy in each interaction, we find that IceCube can probe the dark matter masses up to one gram. 
\end{abstract}

\thispagestyle{empty}  
\newpage  
  
\setcounter{page}{1}  


\section{Introduction}\label{sec:Introduction}

Just like ordinary matter that mainly appears in composite states at a macroscopic scale, dark matter may also have its own interactions and form a composite state that hides its particle identity from observations at a macroscopic scale. Macroscopic Dark Matter (MDM), made of many constituent dark sector particles, is therefore one of compelling paradigms for understanding the particle nature of dark matter. On the particle physics side, there are many models that produce different kinds of MDM: a nontopological extended state that is made of fermions~\cite{Witten:1984rs,Liang:2016tqc,Bai:2018dxf,Hong:2020est,Gross:2021qgx} or bosons~\cite{Rosen:1968mfz,Friedberg:1976me,Coleman:1985ki,Kusenko:1997si,Ponton:2019hux} (see Refs.~\cite{Lee:1991ax,Nugaev:2019vru} for reviews); bound states of asymmetric dark matter particles~\cite{Wise:2014jva,Gresham:2017zqi,Gresham:2017cvl,Coskuner:2018are}; other dark matter bound states including dark blobs~\cite{Grabowska:2018lnd}. On the cosmological side, MDM can be formed via phase transitions~\cite{Witten:1984rs,Bai:2018dxf,Hong:2020est,Gross:2021qgx}, solitosynthesis~\cite{Frieman:1988ut,Griest:1989bq} and dark nucleosynthesis~\cite{Krnjaic:2014xza,Hardy:2014mqa,Redi:2018muu,Mahbubani:2020knq}. Because of MDM states are composite, their masses can be above the Planck mass scale, while their geometric size can also reach the everyday life scale of meters, depending on the energy density of the object (see Ref.~\cite{Jacobs:2014yca} for a review). 

For heavy MDM, detection can be separated into two general approaches: direct or indirect detection. Using a dark matter local energy density of $\rho_{\rm DM} \approx 0.4\,\mbox{GeV}/\mbox{cm}^3$ and averaged dark matter speed of $v_{\rm DM} \approx 10^{-3}\,c$ with $c$ as the speed of light, one anticipates $\mathcal{O}(1)$ encounter events for the dark matter mass of $M_{\rm X} = 1\,\mbox{g} = 5.6\times 10^{23}\,\mbox{GeV}$ with a detector size of kilometer and one-year runtime. For heavier dark matter masses, above one gram, indirect approaches using astrophysical objects to search for MDM become favorable (see Refs.~\cite{Das:2021drz,Bramante:2021dyx} for example). For a lighter MDM, a large terrestrial detector could be ideal to directly detect dark matter. Additional model assumptions about how MDM interacts with the Standard Model (SM) particles are important for determining both the interaction rate and the signal properties. For the simplest case with an elastic scattering of MDM off nuclei in the detector, the deposited energy is bounded by the kinetic energy of the two-body system $\mathcal{O}(10\,\mbox{keV})$. For a large cross section (usually assumed to be a geometric one), MDM can have a multi-hit signature in a detector. One could use the summed energy of those hits or other track-like signal characteristic to pass the trigger requirement or to distinguish signal from backgrounds~\cite{Bramante:2018qbc,Bramante:2018tos}. Some neutrino detectors with a large volume and low energy threshold can search for MDM. For instance, Borexino and JUNO that have a low threshold energy and could detect some multi-hit scattering 
events~\cite{Ponton:2019hux}. 

Other than signals from elastic scattering processes, there is another class of signatures from inelastic scattering processes that is very generic for MDM models and could be adopted to search for MDM. Using the electroweak symmetric dark matter ball (EWS-DMB) for example~\cite{Ponton:2019hux}, two of the authors of this paper have studied the radiative capture process: $\mbox{MDM} + \ce{^A_Z N} \rightarrow \mbox{bound states} + \gamma$~\cite{Bai:2019ogh}. Because of electroweak symmetry restoration inside the ball, the nucleus has a slightly smaller mass if they are inside the dark matter ball than outside, effectively forming a potential well for a nucleus that is inside. MDM can thus capture a nucleus in a detector to form a bound state and release the binding energy into photons using the electromagnetic interaction of the nucleus. The summed photon energy for each interaction is around $A\times 0.25~\mbox{GeV}$~\cite{Bai:2019ogh} with $A$ as the atomic number of the nucleus. The energy released from this inelastic process is in general much larger than the one from elastic scattering and could be accessible to some experiments with a higher energy threshold, such as IceCube. This type of signature is similar to the induced nucleon decay by a Grand Unified Theory (GUT) monopole via the Callan-Rubakov effects~\cite{Callan:1982ac,Rubakov:1982fp}, where a nucleus is captured by the GUT monopole and decay into leptons and mesons. Other than the EWS-DMB model, anti-quark nuggets and Q-balls with an anti-baryon number could annihilate a nucleus and deposit even more energy~\cite{Kasuya:2015uka}. 

Given the fact that the IceCube, with a kilometer-scale detector size, is the largest neutrino and particle physics detector, we demonstrate in this paper that the IceCube detector could be adopted to search for macroscopic dark matter. In fact, it could be the best detector for exploring heavier dark matter masses. In this work, we perform a detailed study of signal and backgrounds to estimate the discovery potential of MDM at IceCube. For the signal events, we will try to keep as much model independence as possible and introduce only two model parameters: the velocity-independent mean free path $\lambdaX$ (related to the inelastic scattering cross section) and the deposited energy $\EX$ (carried away in electromagnetic cascades) for each interaction. The distance between interaction points is sampled from a Poisson distribution with mean $\lambdaX$. For each deposit, we determine the flux of Cherenkov photons on the modules of the detector, which is proportional to $\EX$. Given the similarity to the slow-moving monopole model, we follow closely to the existing analysis by the IceCube collaboration~\cite{IceCube:2014xnp} where a special trigger, the slow particle trigger, is implemented to tag the signature for a slow-moving object (a velocity around $10^{-3}$ of the speed of light). Other than extending signal model parameter space to a larger one with different $\lambdaX$ and $\EX$ beyond the monopole case, we also explore and identify new variables to improve the search sensitivity. In this way, our study could be also useful for searches of non-relativistic GUT monopoles at IceCube. 

The rest of this paper is organized as follows. In Section~\ref{ssec:detector}, we briefly describe some properties of the IceCube and DeepCore detectors. We then discuss the slow particle trigger in Section~\ref{ssec:trigger}, which is important for recording the MDM signal events.  Section~\ref{ssec:simulation} describes the signal and background simulation and two different main cuts to distinguish signal from background. The limits on the dark matter flux and the inelastic cross section are presented in Section~\ref{sec:limits}. We discuss the potential for using the full IceCube detector and summarize this paper in Section~\ref{sec:conclusion}. 

\section{MDM at the IceCube (DeepCore) Detector}\label{sec:detection}

\subsection{IceCube Detector}\label{ssec:detector}
IceCube (IC) Neutrino Observatory under the Antarctic ice has the largest neutrino detector, with a volume over a cubic kilometer. The current IC86/DC configuration consists of 86 strings located at a distance of 1450 meters below the ice surface. Out of these 86 strings, 78 of them, known as IceCube strings, are separated by a horizontal distance of 125 m on average and a horizontal spread of around 1 km in total. Each IC string consists of 60 digital optical modules (DOM) separated by a vertical distance of 17 m, leading to a total height of around 1 km; thus, the IceCube detector is a $1 \,\text{km}^{3}$ detector. 
The remaining 8 strings located at the center of the detector, combined with the 7 IC strings, define the DeepCore (DC) region. In each string, there are a total of 50 such DOMs with a vertical separation of 7 m. The DC region is located between depths of 2100 m to 2450 m with the averaged inter-string separation of DC strings of 72 m. These closely packed DC strings reduce the energy threshold of neutrino detection to be around 10 GeV, compared to the 100 GeV threshold for IC strings~\cite{IceCube:2011ucd}.  

A DOM, with a radius of 16.5 cm, is a glass pressure vessel containing a Hamamatsu photomultiplier tube (PMT) needed to detect Cherenkov light. The center 8 strings in the DC region have higher quantum efficiency (HQE) DOMs, thus reducing the threshold energy of detection. Neutrinos or other particles interacting inside the detector region, producing either electromagnetic (EM) cascade or muon track, give rise to Cherenkov photons. These photons travel in ice and reach the DOMs to produce photoelectrons. A signal passing a threshold of 0.25 photo-electrons (PE) is recorded~\cite{IceCube:2020nwx} and is known as a hit. Two hits recorded on neighbor or next-to-nearest neighbor DOMs on the same string are labeled as {\it high local coincidences} (HLC) if the time difference between the hits is less than $1\,\mu s$. HLC hits act as a basis for any trigger that is constructed for the IceCube detector \cite{IceCube:2008qbc}.

\subsection{Slow Particle Trigger} \label{ssec:trigger}
 
\begin{figure}
	\centering
	\includegraphics[width=0.6\linewidth]{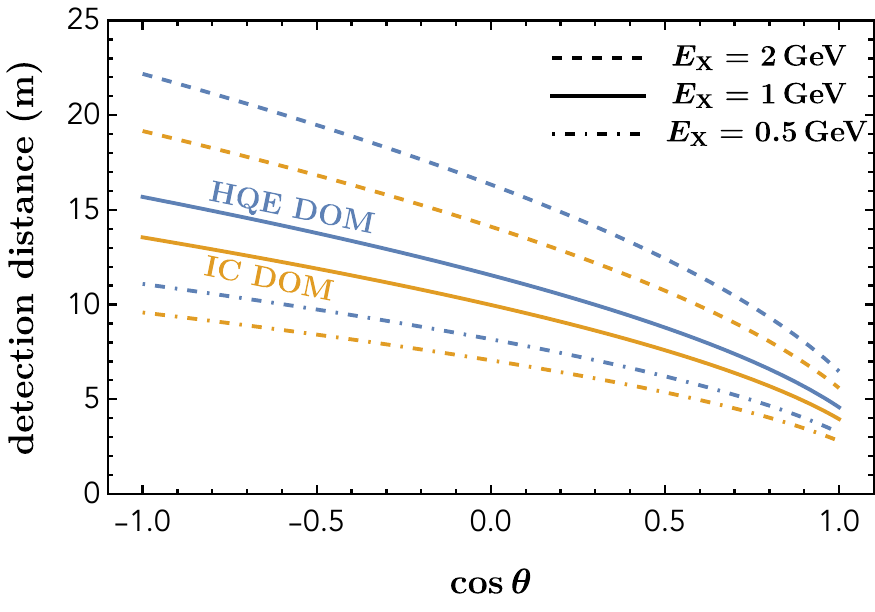}
	\caption{The maximum detection distance (in meters) allowed between an interaction vertex and a DOM as a function of cosine of the angle between the string and the line from the DOM to the vertex~\cite{IceCube:2013llx}. $\EX$ is the deposited energy at each dark matter interaction point. Higher quantum efficiency (HQE) DOMs have a lower threshold energy of detection. 
	}\label{fig.maxdist}
\end{figure}

We now describe the properties of signal to understand the trigger and the relevant cuts. The signal is described by a mean free path along the track ($\lambdaX$) and the energy release in photons ($\EX$) at each interaction point. The released energy is converted to EM cascade which eventually produce Cherenkov photons; the number of resulting Cherenkov photons, in the wavelength range of 300\,nm to 650\,nm, are $\approx 2 \times 10^{5}\,(\EX/\text{GeV})$ and can be approximated as traveling on spherical wavefront centered at the interaction point~\cite{IceCube:2013dkx}.~\footnote{Note that for the parameter space considered in this paper, we do not need to worry about overburden effects reducing the velocity of MDM or completely stopping the MDM before reaching the detector.} Taking into account the quantum efficiency, angular efficiency, and PE threshold, one can determine the maximum detection distance between the interaction vertex and the DOM to generate a hit in the DOM, as a function of the zenith angle. The distributions with different $\EX$ for both IC and HQE DOMs are shown in Fig.~\ref{fig.maxdist}. The effects of scattering and absorption are taken into account by an analytical fit given in Ref.~\cite{IceCube:2013dkx}. Effects of scattering near the DOM are taken into account by a Hole-ice model with an effective modification to the angular efficiency of the DOM~\cite{IceCube:2013llx}. 

\begin{figure}[th!]
	\centering
	\includegraphics[width=1.0\linewidth]{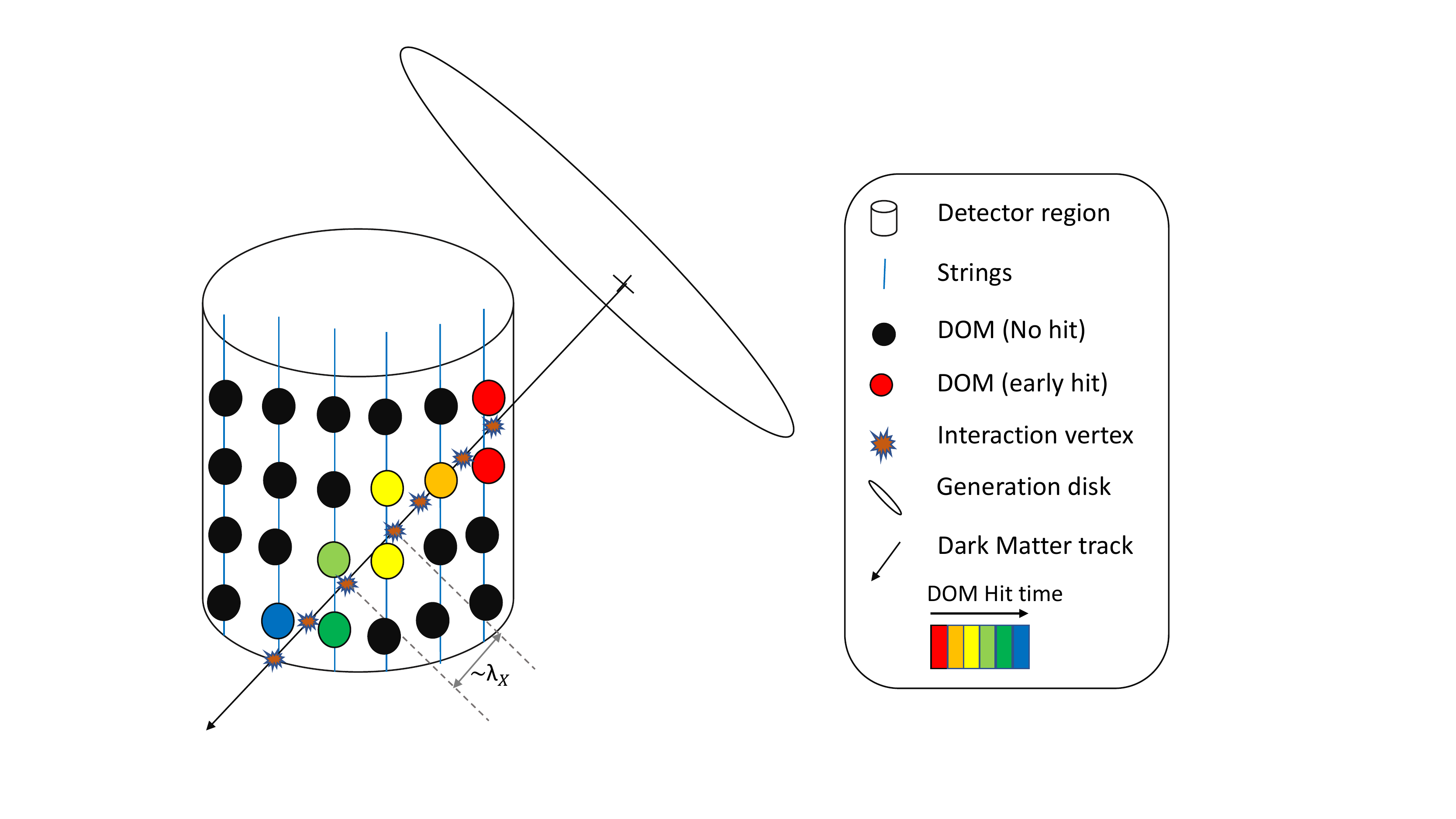}
	\caption{Schematic diagram showing signal simulation and detection at the IceCube detector. The dark matter track (shown by the black arrow) is determined by a randomly chosen starting position on the generation disk with the velocity direction perpendicular to the disk. Interaction vertices are Poisson distributed along the track with a mean separation of the interaction length $\lambdaX$. Each interaction has dark matter inelastically scatter off a nucleus and generate electromagnetic cascades to Cherenkov light. The DOMs that register a hit are shown by the colored circles, with the red colored ones as the earliest hits, while blue ones as the last hits. } \label{fig:signaldetect}
\end{figure}

To simulate the signal events and following Ref.~\cite{IceCube:2014xnp}, we consider a disk placed at a fixed distance from the center of the detector with a schematic plot in Fig.~\ref{fig:signaldetect}. Dark matter objects are randomly placed on the disk with the velocity direction perpendicular to the disk and incoming towards the detector. The disk is rotated randomly around the detector to generate an isotropic flux of dark matter.  Interaction points are Poisson distributed with a mean of $\lambdaX$ on the dark matter track. Given an $\EX$, an appropriate number of Cherenkov photons are generated at each interaction point and approximated to travel on a spherical wavefront. Given the distance between an interaction vertex and a DOM, one can determine the amount of PE charge deposited on a DOM and check whether such an interaction would generate a hit at a given DOM. For an averaged dark matter velocity of $\langle v_{\rm X} \rangle \approx 300\,\text{km}/\text{s}$, the time at each interaction point and thus the time of a hit at a DOM is determined. The time information of hits will be used to distinguish signal from background events. The HLC hits are located near the dark matter track and distributed over a time scale of $\mathcal{O}(1\,\mbox{ms})$ (time required for dark matter to cross the whole detector). The time difference between the HLC hits is proportional to the distance between HLC hits, as expected from dark matter traveling at a constant speed.  

The backgrounds for the signal event contain Poisson noise in the DOM due to radioactive decays and atmospheric muons leaving a track of HLC hits. HLC hits produced by Poisson noise are randomly distributed over the detector volume with no correlation between the HLC hit time and distance. On the other hand, since muons travel at relativistic speeds, the HLC hits generated from them are distributed over a few microsecond time scales (the time for muons to cross the whole detector).  These distinguishing characteristics between signal and backgrounds are used to design a slow particle trigger (SLOP trigger)~\cite{IceCube:2014xnp}, which we will use for the analysis in this paper. 

\begin{figure}[t!]
	\centering
	\includegraphics[width=0.45\linewidth]{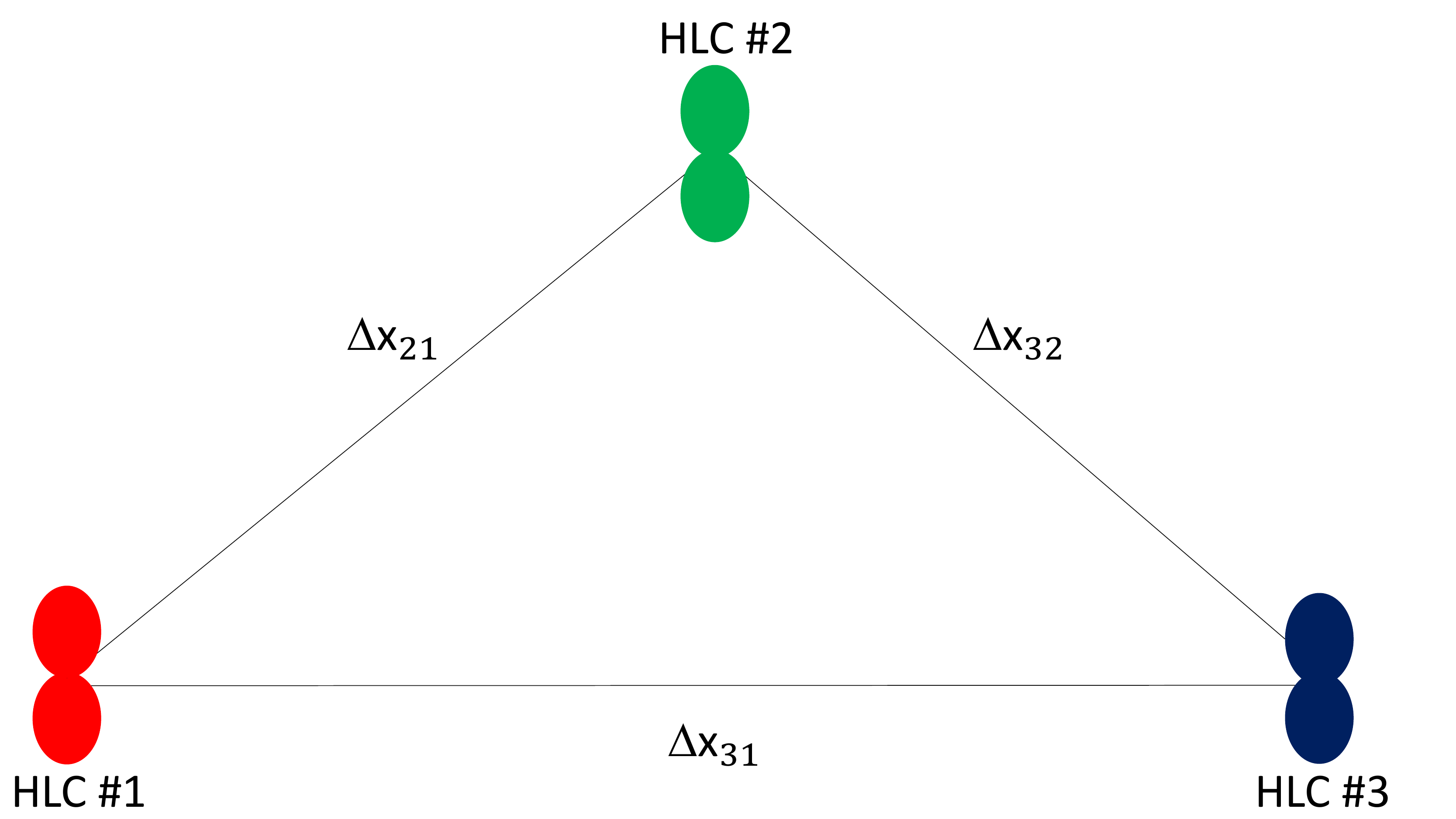}
	\hspace{6mm}
	\includegraphics[width=0.45\linewidth]{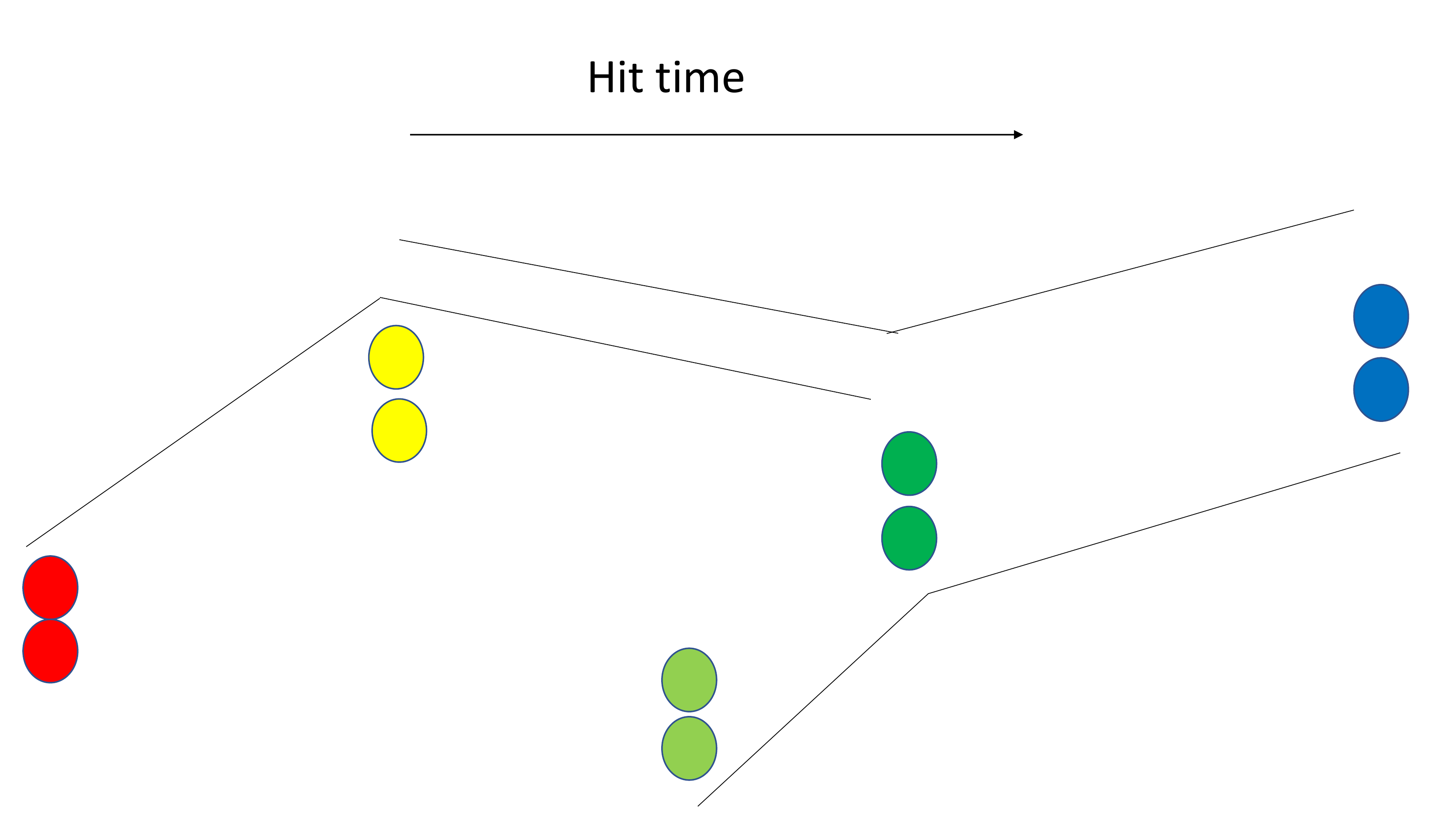}
	\caption{Left panel: example of triplet that is made out of three HLC hits. The distances between HLC hits and the time differences are used to calculate the parameters $\Delta d$ and $v_{\text{rel}}$. Right panel: a set of triplets overlapping in time is shown. HLC hits are time ordered, with time increasing from left to right. All possible triplet combinations that satisfy the cuts on $\Delta d$ and $v_{\text{rel}}$ are counted towards $n_{\text{triplet}}$. } \label{fig:triplets}
\end{figure}

To reduce the atmospheric muon background, all HLC hits with a time difference less than $\Delta t < t_{\text{proximity}} = 2.5\,\mu s$ are removed. The remaining HLC hits are used to create all possible time ordered combination of three HLC pairs, called {\it triplets}. The time difference between any two HLC hits in a triplet is restricted to $[t_{\text{min}}, t_{\text{max}}]$ with $t_{\text{min}} = 0$ and $t_{\text{max}} = 500\,\mu s$. These two constraints can be combined into one constraint: $\Delta t_{\text{HLC}} \in (2.5 , 500)\,\mu s$. HLCs in the triplet created by the signal lie along a line and have time difference between them consistent with a constant speed. To further match the signal event topology, two more parameters $\Delta d $ and $v_{\text{rel}}$ are introduced. The parameter $\Delta d = \Delta x_{21} +\Delta x_{32} - \Delta x_{31}$, with $\Delta x_{i j}$ as the distance between HLC hits $i$ and $j$ in a given triplet (see the left panel of Fig.~\ref{fig:triplets}). A value of $\Delta d = 0$ corresponds to a straight line. The SLOP trigger requires $\Delta d \leq 100\,\mbox{m}$. The second parameter $v_{\text{rel}}$ is defined as 
\beqa
\label{eq:vrel}
v_{\text{rel}} = \frac{\Big{|}\frac{1}{v_{21}} - \frac{1}{v_{32}}  \Big{|}}{\frac{1}{3}\cdot \Big{(}\frac{1}{v_{21}} + \frac{1}{v_{32}} + \frac{1}{v_{31}}\Big{)}} ~,
\eeqa
where $v_{i j} = \Delta d_{i j}/ \Delta t_{i j}$ is speed between $i$'th and $j$'th HLC pair in a triplet, with $\Delta t_{i j}$ being the time difference between corresponding HLC hits. Note that  $v_{\text{rel}}$ is dimensionless and the specific form is chosen to make it independent of the overall dark matter speed. A value of $v_{\text{rel}} = 0$ corresponds to a dark matter track with a constant speed. The SLOP trigger requirement is that $v_{\text{rel}} \leq 0.5$. All the triplets not satisfying the trigger cuts on these parameters are removed from the list of triplets. Out of remaining triplets, there is a trigger condition requiring a set of triplets overlapping in time (see the right panel of Fig.~\ref{fig:triplets}) containing $n_{\text{triplet}}\geq 3$ triplets. The event duration of a triggered event, that is the time difference between the first HLC hit of the first triplet and the last HLC hit of the last triplet in the time-ordered triplets, is required to be shorter than $5 \,\text{ms}$~\cite{IceCube:2014xnp}.

The SLOP trigger has been operating on the DeepCore detector since May 2011 and on the whole IceCube detector since May 2012~\cite{IceCube:2014xnp}. The trigger rate is 2.1\,Hz for the DC detector and 12\,Hz for the IC detector (with slightly different trigger cuts for IC)~\cite{Kelley:2015ncf}. Two days of experimental data were used to study the background characteristics assuming that such a small amount of data is unlikely to contain any signal events. The triplet distribution can be fitted well with a random noise model as described in Ref.~\cite{IceCube:2014xnp}. 
According to this model, the probability to obtain $n$ triplets is given by 
\beqa
\label{eq:noisemodel}
P(n\,|\, \mu, p) = P_{0}\,\sum_{N=N_{\text{min}}(n)} P_{\mu}(N)\, B\big(n\,|\,n_{\text{max}}(N), p\big) \, ,
\eeqa
where $P_{\mu}(N) $ is a Poisson probability distribution to obtain $N$ HLC hits with $\mu$ being the mean expectation of the number of HLC hits  in a given time window, implying HLC hits arising from random noises follow a Poisson distribution. Given a certain number of HLC hits, the maximum number of possible triplets  is $n_{\text{max}}(N) = \left(\begin{smallmatrix}N\\3\end{smallmatrix}\right)$. Hence, the probability of obtaining $n$ triplets is given by a Binomial distribution $B(n\,|\,n_{\text{max}}(N),p) = \left(\begin{smallmatrix}n_{\text{max}}\\n\end{smallmatrix}\right)p^n (1-p)^{n_{\text{max}}-n}$ with $p$ being the probability of success. The sum in \eqref{eq:noisemodel} starts with $N_{\text{min}}(n)$, the minimum number of HLC hits required to get $n$ triplets. $P_{0}$ is the overall normalization.

\subsection{Signal and Background Simulation and Cuts} \label{ssec:simulation}

\begin{figure}[th!]
	\centering
	\includegraphics[width=0.48\linewidth]{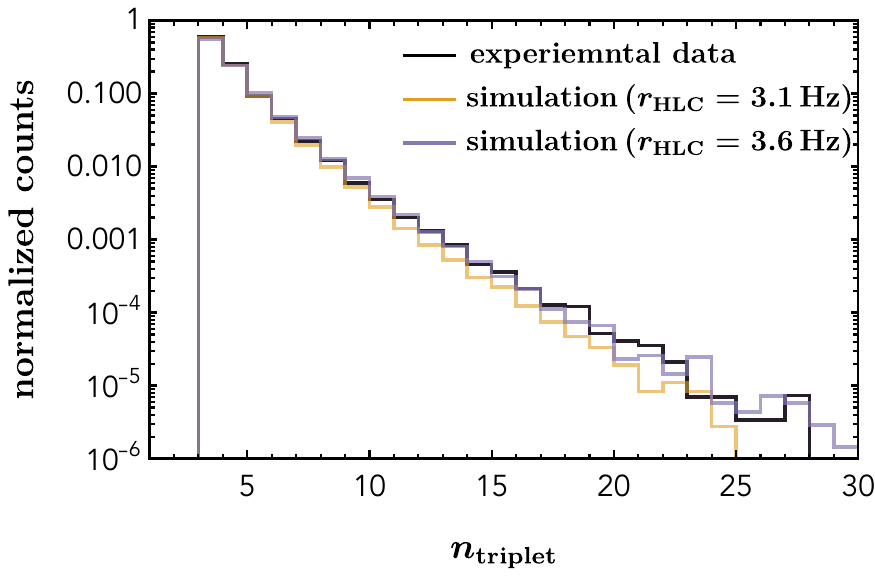}
	\hspace{3 mm}
	\includegraphics[width=0.47\linewidth]{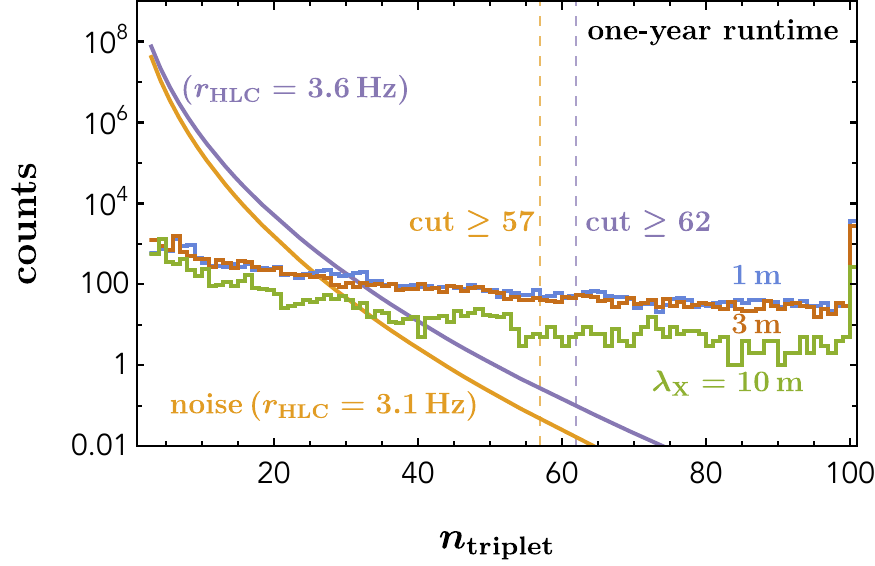}
	\caption{Left panel: the triplet distributions for the two-day experimental data (black curve)
	and the simulated noise background with the HLC rates of $r_{\text{HLC}}=3.1\,\text{Hz}$ (orange curve) and $3.6\,\text{Hz}$ (violet).	Right panel: the noise background and dark matter signal count distributions in $n_{\rm triplet}$ for one-year runtime and $v_{\rm X} = 300$\,km/s. The noise backgrounds follow the fitted function in \eqref{eq:noisemodel} with $P_{0} = (4.82, 5.25) \times 10^{10}$, $\mu=(1.05,1.2)$, and $p =(0.15,0.15)$ for $r_{\text{HLC}} = (3.1, 3.6)\, \text{Hz}$.  A triplet cut with $n_\text{triplet} \geq (57, 62)$ removes almost all background events and will be used to distinguish signal from background.  }\label{fig:noisecompare}
\end{figure}

To simulate the random noise background, we assign all the DOMs with a fixed HLC rate obtained by computing the SLOP trigger rate and matching it with the observed value. For DC, we have found that a HLC rate ($r_{\text{HLC}}$) around $3.1 \,\text{Hz}$ can provide a SLOP trigger rate of $2.1 \,\text{Hz}$ (the measured one~\cite{IceCube:2014xnp}).  In the left panel of Fig.~\ref{fig:noisecompare}, we compare the 2-day SLOP trigger data (black curve) with the simulated backgrounds with $r_{\text{HLC}}=3.1 \,\text{Hz}$ (orange curve) and a slightly higher rate of 3.6\,Hz (violet curve), which corresponds to a SLOP trigger rate of 4.0\,Hz, but has a better fit to the triplet distribution. We keep both choices to demonstrate that our final results are insensitive to the choice of $r_{\text{HLC}}$. The triplet distributions of our simulation for both choices of $r_{\text{HLC}}$ agree reasonably well with the one from data.  We can then use it to obtain the parameters of the noise background model in \eqref{eq:noisemodel}. Normalizing to one-year runtime, we find that $\mu = 1.05$, $p = 0.15$, and $P_{0} = 4.82 \times 10^{10}$ for $r_{\text{HLC}}=3.1~\text{Hz}$, and $\mu = 1.2$, $p = 0.15$, and $P_{0} = 5.25 \times 10^{10}$ for $r_{\text{HLC}}=3.6\,\text{Hz}$. In the right panel of Fig.~\ref{fig:noisecompare}, we compare the triplet distributions for noise background with $r_{\text{HLC}}=3.1\, \text{Hz}$ (orange curve) and $3.6\, \text{Hz}$ (violet curve) and the signals with $\EX = 1$~GeV and different interaction lengths $\lambdaX=1\,\text{m}$ (blue curve), $\lambdaX = 3\,\text{m}$ (brown curve), and $\lambdaX =10\,\text{m}$ (green curve).
From those distributions, one can see that the number of triplets, $n_\text{triplet}$, can be used to distinguish the noise background from the signal (as also demonstrated in Ref.~\cite{IceCube:2014xnp}). As expected, the dark matter signal events have a distribution populating larger $n_\text{triplet}$ compared to the noise background distribution. The signal event counts in $n_\text{triplet}$ increases from the longer interaction length $\lambdaX = 10$\,m to the shorter one with $\lambdaX = 3$\,m, but saturates afterwards. This saturation is because the dead time (the time period during which the DOM cannot receive a second HLC hit)  of the DOM is around 6.4\,$\mu$s for the HLC hit~\cite{IceCube:2008qbc}, which approximately matches the dark matter mean free time for $\lambdaX = 3$\,m and $v_{\rm X} = 300$\,km/s. Choosing a cut of $n_\text{triplet} \geq 57$ (the one used in Ref.~\cite{IceCube:2014xnp}) for $r_{\text{HLC}}=3.1\,\text{Hz}$, the search becomes background-free (less than 0.5 events for one year data). For $r_{\text{HLC}}=3.6\,\text{Hz}$, one must  impose a slightly more stringent cut with $n_\text{triplet} \geq 62$ to be background free, but the change in signal efficiencies is only a few percent compared to the $n_\text{triplet} \geq 57$ cut.

\begin{figure}
	\centering
	\includegraphics[width=0.47\linewidth]{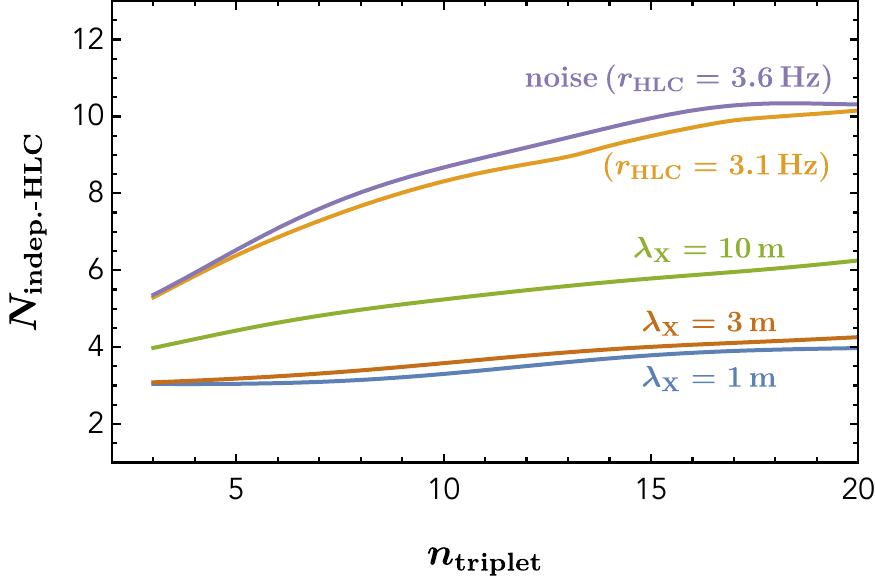}
	\hspace{3 mm}
	\includegraphics[width=0.48\linewidth]{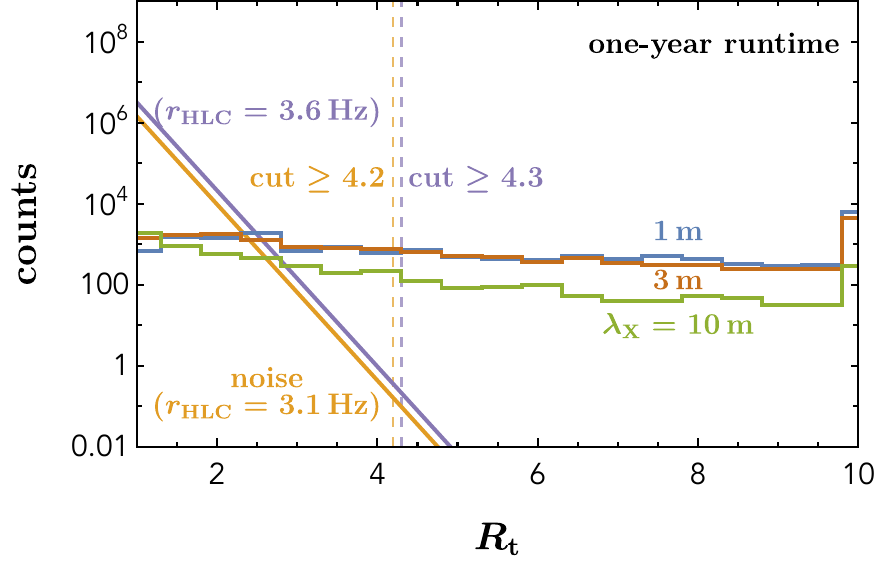}
	\caption{Left panel: the number of triplets vs.~the average number of independent HLC positions for both the noise background and signal with different interaction lengths. A 2-day runtime has been used for the simulation. Right panel: the counts as a function of the ratio variable defined in \eqref{eq:ratio-define} for both the noise background and the signal. A one-year runtime has been used here. The noise background can be well fit by a function $f(x) = a \, e^{-b\,x}$ with $a = (2.1, 4.7) \times 10^{8}$ and $b = (5, 5)$ for $r_{\text{HLC}}=(3.1, 3.6) \,\text{Hz}$. The $R_{\rm t}$ cut of $\geq (4.2, 4.3)$ can approximately remove all background and will be used in later analysis to select signal from background. For both panels, $\EX=1$\,GeV and $v_{\rm X} = 300$\,km/s. 
	 }\label{fig:ratio}
\end{figure}

Another distinguishing characteristic between the signal and the noise background is the number of independent HLC positions needed to make a certain number of overlapping triplets. The corresponding distributions for signals and backgrounds are shown in the left panel of Fig.~\ref{fig:ratio}. For both signal and background, more independent HLC positions are needed to produce a certain number of triplets. Note that independent HLC positions are different from independent HLC hits, since a single DOM undergoing HLC hit at one time can also give an HLC hit at another (later) time. These two HLC hits at the same DOM can then contribute to different triplets, although they have the same HLC position. It is common for the signal to produce multiple HLC hits in the same DOM. Once a dark matter object passes in the vicinity of a DOM, the signals from several different interaction points, separated on average by a distance $\lambdaX$, can deposit enough light in the DOM to give multiple HLC hits (even after taking into account the HLC $6.4\,\mu\mbox{s}$ dead time~\cite{IceCube:2008qbc}).  This observation, combined with the fact that the signal track-like events have a higher probability of forming a triplet compared to the noise background events that are randomly distributed over the detector, explains the noise vs. signal trend in the left panel of Fig.~\ref{fig:ratio}. 

\begin{table}[hb!]
	\centering
	\renewcommand{\arraystretch}{1.4}
	{\footnotesize
		\begin{tabular}{|l | l | l | l | l | l | l | l |}
			\hline
			\hline
			\multirow{2}{*}{\diagbox[innerwidth=4cm]{Cut}{}}
			&
			\multicolumn{2}{c}{Background efficiency ($r_{\text{HLC}}$)} 
			&
			\multicolumn{5}{|c|}{Signal efficiency ($\lambdaX= 1 \text{m}$) ($\EX$)}  \\\cline{2-8}
			& $3.6~\text{Hz}$ & $3.1~\text{Hz}$ & 4 GeV & 2 GeV & 1 GeV & 0.5 GeV & 0.25 GeV \\\hline\hline
			$n_{\text{DOM}}\geq 1$ & 1.000 & 1.000 & 0.497 & 0.410 & 0.328 & 0.238 & 0.163 \\ \hline
			$n_{\text{HLC}}\geq 1$ & 1.000 & 1.000 & 0.449 & 0.331 & 0.204 & 0.131 & 0.063 \\ \hline
			$\Delta t_{\text{HLC}}\in (2.5 ,500)\,\mu s$ &  0.764 & 0.649 & 0.307 & 0.191 & 0.0905 & 0.0236 & 0.00354 \\ \hline
			$\Delta d\leq 100 \, \text{m}$, $v_{\text{rel}}\leq 0.5$ & 0.217 & 0.151 & 0.302 & 0.180 & 0.0898 & 0.0234 & 0.00354 		\\ \hline
			$n_\text{triplet} \geq 3$ & 0.0200 & 0.0105 & 0.299 & 0.174 & 0.0865 & 0.0212 & 0.00161 \\ \hline 	\hline
			$n_\text{triplet} \geq 57$  & $2.17 \times 10^{-10}$ & $3.70 \times 10^{-11}$ & 0.235 & 0.0822 & 0.0229 &  0.00298 & 0.000113 \\ 
			\hline 
			$R_{\text{t}}\geq 4.2$ & $1.43 \times 10^{-10}$ & $6.48 \times 10^{-11}$ & 0.275 & 0.126 & 0.0465 & 0.00663 & 0.000375 \\ \hline
			\hline
		\end{tabular}
	}
	\caption{
	Background and signal efficiencies after different cuts. 
The fist two rows of cuts, $n_{\text{DOM}}\geq 1$ and $n_{\text{HLC}}\geq 1$, imply at least one DOM hit and one HLC hit in the entire DC detector in the event time window. The next three rows are the SLOP-trigger cuts discussed in the text. The last two rows are independent cuts on $n_\text{triplet}$ and $R_{\text{t}}$ variables to have the number of background events less than one for one-year runtime. The dark matter velocity is fixed to have $v_{\rm X} = 300\,\mbox{km}/\mbox{s}$. For 10-year runtime, we will use $n_\text{triplet} \geq 68$ and  $R_{\text{t}}\geq 4.7$. } \label{tab:eff}
\end{table}

To make use of this distinguishing characteristic between the signal and the noise background, we introduce a new variable, $R_{\rm t}$, defined as the ratio of number of overlapping triplets and independent HLC positions needed to make those triplets: 
\beqa
\label{eq:ratio-define}
R_{\rm t} \equiv \frac{n_{\rm triplet}}{N_{\rm indep.-HLC}}~.
\eeqa
In the right panel of Fig.~\ref{fig:ratio}, we compare the $R_{\rm t}$ distributions for the noise background with $r_{\text{HLC}}= 3.1\,\text{Hz}$ (orange curve) and $3.6\,\text{Hz}$ (violet curve) and the signals with $\lambdaX = 1 \, \text{m}$ (blue curve), $\lambdaX = 3 \, \text{m}$ (brown curve), $\lambdaX = 10 \, \text{m}$ (green curve), all with $\EX = 1$\,GeV. The tail distribution of the noise background is fit to an exponential function with $f(x) = a \, e^{-b\,x}$ with $a = (2.1, 4.7) \times 10^{8}$ and $ b = (5, 5)$ for $r_{\text{HLC}}=(3.1, 3.6) \,\text{Hz}$. We have found that we can replace the $n_\text{triplet}$ cut by a cut on the new variable $R_{\rm t}$ to increase the ratio of signal over background. Specifically, the ratio cut with $R_{\rm t} \geq (4.2, 4.3)$ for $r_{\text{HLC}}=(3.1, 3.6)\, \text{Hz}$ can make background free while keep signal efficiencies  higher than using the $n_\text{triplet}$ cut. Since the signal efficiency merely changes by a few percent by using a more stringent cut for the larger noise rate, we will just impose the ratio cut $R_{\rm t} \geq 4.2$ to obtain the limits on the dark matter flux. 

To summarize effects of various cuts, we show both background and signal efficiencies after these cuts in Table~\ref{tab:eff}. The first two rows of cut: $N_{\text{DOM}}\geq1$ and $N_{\text{HLC}} \geq 1$ imply at least one DOM hit and one HLC hit in the entire detector for a given event. For the noise background, these values saturate to be one for one-year runtime, while the signal efficiency decreases with decreasing energy. The next three rows are comprised of the SLOP trigger cuts discussed above. As expected, the reduction on the noise background efficiency is much larger compared to the reductions on the signal efficiencies. The  last two rows are the cuts on $n_\text{triplet}$ and $R_{\rm t}$ to obtain a background free search. One can clearly see that the variable $R_{\rm t}$ more efficiently improves the signal to background ratio than $n_\text{triplet}$.

\section{Limits on Dark Matter Flux and Cross Section}\label{sec:limits}

To evaluate the sensitivity on the dark matter flux, we impose cuts yielding zero expected background events ($n_{b}= 0$) and assume null observation of signal events ($n_{\text{obs}} = 0$). In absence of signal and with zero background, the 90\% confidence level (CL) flux limit is given by  
\beqa
\Phi_{90} = \frac{\overline{\mu}_{90}}{A_{\text{eff}}\cdot T \cdot \Omega} ~ ,
\eeqa
where $\overline{\mu}_{90} = 2.44$~\cite{Feldman:1997qc}, $\Omega$ is the solid angle $4\pi$, $T$ is the total runtime of the detector, and $A_{\text{eff}} = \epsilon \cdot A_{\text{gen}}$ is the effective area with $\epsilon$ as the signal efficient and $A_{\text{gen}}$ is the area of generating disk. For the DC case, we choose the radius of the generating disk as $r_{\text{gen}} = 250 \, \text{m}$ corresponding to $A_{\text{gen}}= \pi \, r_{\text{gen}}^{2} \approx 2 \times 10^{5}\,\mbox{m}^{2}$.

\begin{figure}[t!]
	\centering
	\includegraphics[width=0.47\linewidth]{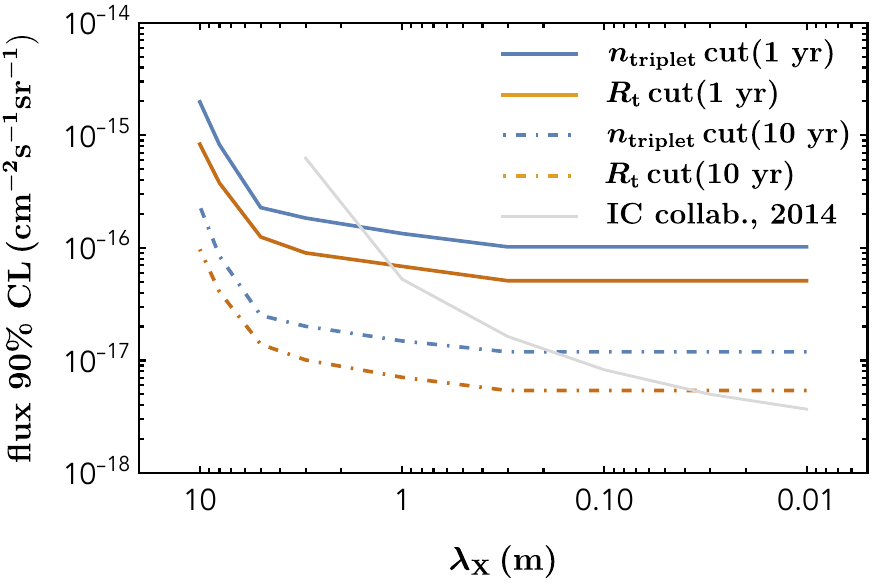}
	\hspace{3mm}
	\includegraphics[width=0.47\linewidth]{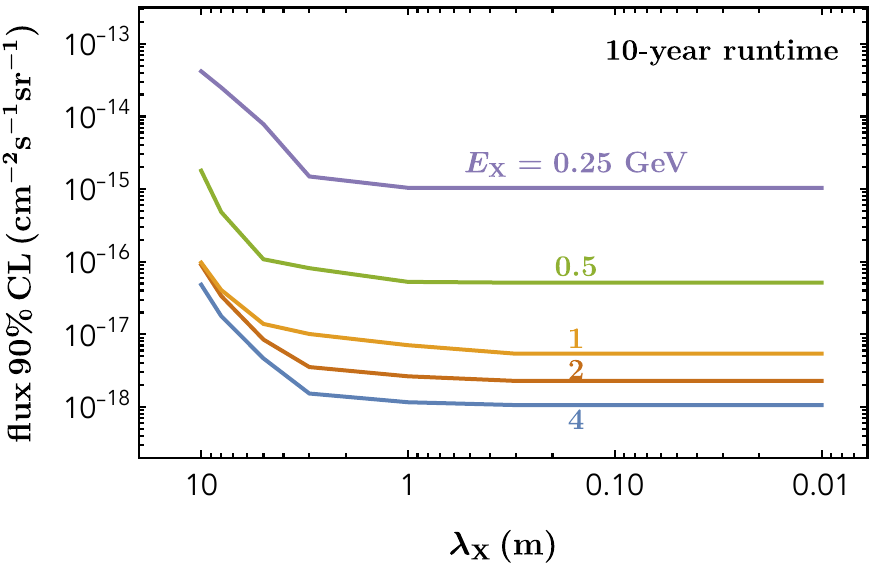}
	\caption{Left panel: projected limits on macroscopic dark matter flux at 90\% CL as a function of the interaction length $\lambdaX$. The DeepCore detector is used in the analysis. The dark matter particles are assumed to have a fixed interaction-deposited energy $\EX = 1 \, \text{GeV}$ and velocity $v_{\rm X} = 300$\,km/s. Both limits from the triplet cut and the ratio cut are shown for comparison. We also show the existing IC86/DC (questionable) limits from the IceCube collaboration in Ref.~\cite{IceCube:2014xnp} (see text for more discussion), which is based on one-year runtime data. Right panel: the same as the left panel but for different $\EX$ using the ratio $R_{\rm t}$ cut with 10-year runtime. 
	}\label{fig:fluxlimit}
\end{figure}

In the left panel of Fig.~\ref{fig:fluxlimit}, we show the projected limits on the dark matter flux for different interaction lengths $\lambdaX$ and fixed $\EX = 1\, \text{GeV}$. The limits using the new ratio variable $R_{\rm t}$ (brown curve) are more than a factor of two stringent compared to the one using the $n_{\rm triplet}$ cut (blue curve). The limits from 10-year runtime (dot-dashed) curve are almost a factor ten more stringent than the limits from 1-year runtime (solid). For the 10-year runtime results, we use the triplet cut of $n_{\rm triplet}\geq 68$ and the ratio cut of $N_{\rm t}\geq 4.7$ to obtain approximately zero expected background. 

The gray curve in left panel of Fig.~\ref{fig:fluxlimit} shows current experimental limits derived by the IceCube Collaboration~\cite{IceCube:2014xnp}. In their analysis, only the cut on the number of triplets is used to obtain limits with one-year data. IceCube's results are then a factor of 10 or 100 stronger than the results of our analysis when reducing $\lambdaX$ from 1 m to 0.1 m or 0.01 m respectively. This disagreement is at odds with our qualitative understanding of the signal generation and detection.   

Even though reducing $\lambdaX$ can increase the number of interaction points, the energy released at each interaction vertex is still 1 GeV (for the monopole case in \cite{IceCube:2014xnp}), which limits the number of DOMs that could register a hit. To be more quantitative, one can calculate the effective area to produce a single hit in the entire detector, which saturates to $66000\,\mbox{m}^{2}$ for the DC detector with $\EX = 1$\,GeV. Any effective area derived after the triplet cut has to be smaller than this area. However, the limits derived in Ref.~\cite{IceCube:2014xnp} do not follow the expected behavior. The correct limits should saturate when $\lambdaX$ is around one meter, which is the case for the blue and brown curves in the left panel of Fig.~\ref{fig:fluxlimit}. The signal efficiency decreases for  $\lambdaX$ larger than a few meters because these interaction lengths are comparable to the maximally allowed distance between an interaction vertex and a DOM to detect a hit in the DOM (see Fig.~\ref{fig.maxdist}) and to the inter-DOM separation of the DC detector. 

In the right panel of Fig.~\ref{fig:fluxlimit} and to target on different dark matter models, we show limits on the dark matter flux as a function of $\lambdaX$ for different interaction-deposited energies. In this plot, we have chosen a 10-year runtime. One can clearly see that the sensitivity drops fast once the deposited energy is below 1 GeV. Also note that there is a universal feature for all curves to have a weaker limit for a large interaction length $\lambdaX$.

\begin{figure}[t!]
	\centering
	\includegraphics[width=0.55\linewidth]{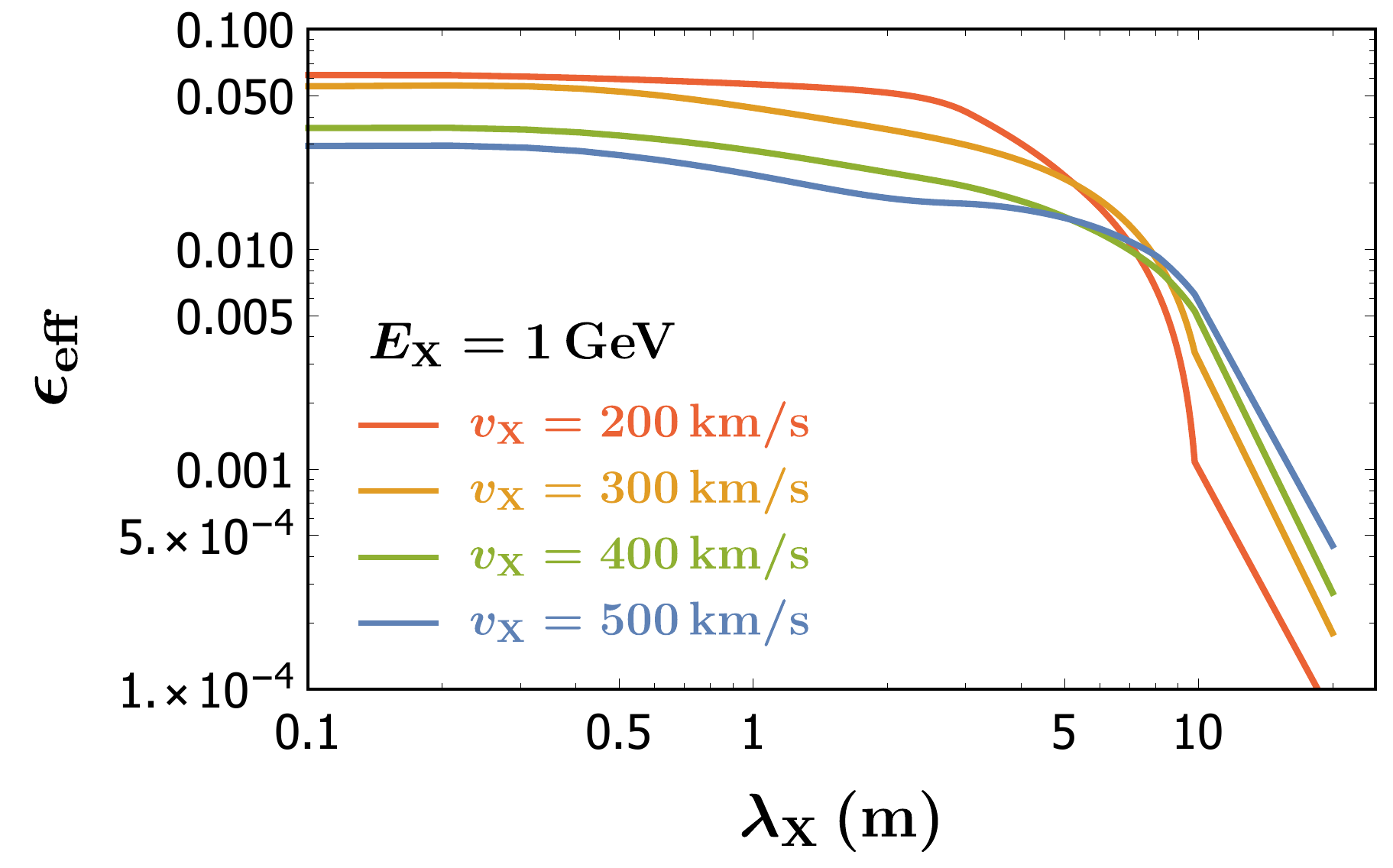}
	\caption{Signal efficiency $\epsilon_{\text{eff}}(v_{\rm X},\lambdaX, \EX)$ as a function of 
	$\lambdaX$ for different dark matter velocities $v_{\rm X}$ and fixed $\EX = 1 \, \text{GeV}$. Here, we have used the cut for a 10-year runtime: $R_{\text{t}}\geq 4.7$.
}
	\label{fig:velocity}
\end{figure}

Note that we have restricted our analysis $\lambda_{\rm X} \geq 0.01\,\text{m}$. For the case with a smaller interaction
length below around $10^{-4}$\,m, our analysis method breaks down, as the mean free time for the MDM would be smaller than the time resolution of the DOM, $\mathcal{O}(\text{ns})$~\cite{IceCube:2008qbc}. In this case, we would need to sum up the energy released from multiple interaction vertices,  thus effectively leading to a larger $E_{\text{X}}$. The limits for the $\lambda_{\rm X} \ll 10^{-4}\, \text{m}$ case could be better than the one for $\lambda_{\rm X} = 0.01\, \text{m}$. 

Up to now, we have fixed the dark matter velocity as $v_{\rm X} = 300 \, \text{km}/\text{s}$. To convert the limits on the dark matter flux into constraints on the dark matter mass and interaction cross section, one need to know the signal efficiencies for different dark matter velocities, which are shown in Fig.~\ref{fig:velocity}. In this plot, some peculiar characteristics can be observed. First, for small $\lambdaX$, the signal efficiency is higher for smaller velocities, while for large $\lambdaX$ the signal efficiency is higher for larger velocities. The small $\lambdaX$ behavior can be explained by the fact that we expect continuous hits in the DOMs but this is limited by the dead time of the DOMs. For larger velocities, more hits would fall in the dead time of an earlier HLC hit, giving a smaller number of HLC hits and a smaller number of triplets, and hence a lower signal efficiency. The behavior for larger $\lambdaX$ can be explained by the SLOP trigger requirement $\Delta t_{\text{HLC}} \leq t_{\text{max}} = 500 \,\mu \text{s}$, which constrains the time difference between the HLC hits of a given triplet. This requirement implies that only HLC hits on strings with a separation distance smaller than $150\, \text{m} \times [v_{\rm X}/(300\,\text{km}/\text{s})]$ can make a triplet. For smaller velocities, this limits the number of triplets and hence the signal efficiency. 

Taking into account the local dark matter velocity distribution, one can obtain a constraint on the dark matter mass and cross section or the interaction length as 
\beqa
\label{eq:masslimit}
M_{\text{X}} &>& 3 \times 10^{24} \, \text{GeV} \,\times\,\Big{(}\frac{\rho_{\text{DM}}}{0.4 \, \text{GeV}/\text{cm}^{3}}\Big{)} \Big{(}\frac{A_{\text{gen}}}{2 \times 10^{5} \, \text{m}^{2}} \Big{)} \Big{(}\frac{T}{10 \, \text{yrs}} \Big{)} \nonumber \\
&& \times 
 \int d v_{\rm X} \, f_{\text{DM}}(v_{\rm X}) \, \epsilon_{\text{eff}}(v_{\rm X}, \lambdaX, \EX)\,\left(\frac{v_{\rm X}}{300 \, \text{km}/\text{s}}\right)   \, ,
\eeqa
where $\rho_{\text{DM}}$ is the local dark matter energy density and $f_{\text{DM}}(v)$ is the dark matter velocity distribution function in the lab frame, which we use the Maxwellian Standard Halo Model~\cite{Necib:2018igl} and have checked that the non-Maxwellian distribution makes a negligible difference. We have ignored the directional dependence of the flux and efficiency, which have small effects for the limits. 

\begin{figure}[t!]
	\centering
	\includegraphics[width=0.6\linewidth]{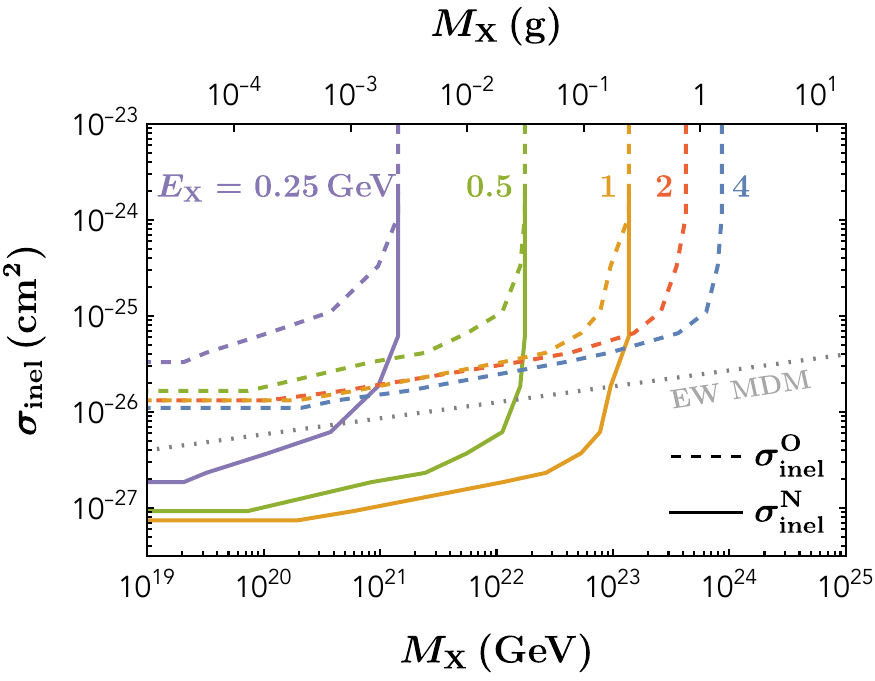}
	\caption{Projected 90\% CL limits on the dark matter mass and inelastic scattering cross section for different deposition energy of each interaction, assuming 10-year runtime of the DeepCore detector. A local energy density of $\rho_{\rm DM} = 0.4\,\mbox{GeV}/\mbox{cm}^3$ is used here. The inelastic cross sections $\sigma^{\text{O}}_{\rm inel}$ and $\sigma^{\text{N}}_{\rm inel}$ are those for capturing an oxygen atom and a nucleon, respectively. The gray dotted line is the radiative capture cross section for a particular MDM model with an electroweak energy density~\cite{Bai:2019ogh}. }\label{fig:cross-section}
\end{figure}

The interaction length $\lambdaX$ is related to the inelastic scattering cross section $\sigma_{\rm inel}$, which depends on whether the individual nucleon or the whole nucleus is bound to or annihilated by dark matter. To be more independent of the detailed dark matter model, we consider two cases: 1) interaction cross section $\sigma_{\rm inel}^{\rm N}$ with an individual nucleon (total of 18 nucleons in one ice molecule), and 2) interaction cross section $\sigma_{\rm inel}^{\rm O}$ with the whole oxygen nucleus (one nucleus per ice molecule). Using the ice density of $\rho_{\text{ice}} \approx 0.9 \, \text{g}/\text{cm}^{3}$~\cite{ice-property}, the relation between $\sigma_{\rm inel}$ and $\lambdaX$ is 
\beqa
\label{eq:cross_sec}
\sigma_{{\rm inel}}^{\text{N}} = 1.85\,\times 10^{-26} \,\mbox{cm}^2\, \times\,\left(\frac{1 \, \text{m}}{\lambdaX} \right) ~,
\qquad 
\sigma_{{\rm inel}}^{\text{O}} = 33 \times 10^{-26} \,\mbox{cm}^2\, \times\,\left(\frac{1 \, \text{m}}{\lambdaX} \right)
 ~.
\eeqa
We have assumed both $\sigma_{\rm inel}$ and $\lambdaX$ to be velocity-independent, which may be not be true for some models, but only brings a small modification for the limit in~\eqref{eq:masslimit}. 

Combining \eqref{eq:cross_sec} and \eqref{eq:masslimit}, we show the projected sensitivity on the dark matter mass and cross section in Fig.~\ref{fig:cross-section} for different deposition energy $\EX$ and 10-year runtime. For a larger value of $\EX$, one has a more stringent limit on the dark matter flux and hence can probe a heavier dark matter mass. For the interaction cross section with a nucleon, the maximum energy deposit is the nucleon mass, approximately 1 GeV, but for dark matter capturing an oxygen nucleus, the released energy could be large, \eg, around 4 GeV for an EWS-DMB~\cite{Ponton:2019hux,Bai:2019ogh}.

\section{Discussion and Conclusions}\label{sec:conclusion}

In the above section, we have only considered the DeepCore detector to derive limits; considering the full IC86/DC detector is not expected to improve the limits that much for $E_{\text{X}} \leq 1 \, \text{GeV}$. For the IC86 detector with IC strings, the inter-DOM separation is 17\,m, which is large compared to the maximum distance allowed between the interaction vertex and the DOM as shown by the orange solid and dot-dashed curve in Fig.~\ref{fig.maxdist}. Thus, for the $E_{\text{X}} \leq 1 \, \text{GeV}$ case, MDM is mainly detected in the DeepCore region giving similar flux limits as shown in Fig.~\ref{fig:fluxlimit}. For $\EX \leq 1$\,GeV, the flux limits can be improved by using a denser detector compared to the DeepCore detector, such as PINGU \cite{IceCube:2016xxt}.  For the case of $E_{\text{X}} \geq 2\, \text{GeV}$ and $4\, \text{GeV}$, we expect the full IC86/DC detector provides a better limit. Based on a similar background model as for DeepCore, \ie, noise being the dominant component, we obtain a factor of 2 and 5 improvement on the flux limits, respectively. To derive reliable limits using the full detector, one also needs a data-driven approach to study the background as was done for the DeepCore case \cite{IceCube:2014xnp}. 

In summary, we have considered a direct detection probe of a class of MDM,
which can either radiatively capture nuclei or induce nucleon decay to release the energy. We have demonstrated that the DeepCore region of the IceCube neutrino detector has sensitivity to probe a significant portion of the parameter space for such models. Our analysis is based on an existing slow particle trigger to collect signal events. We have identified a new variable ($R_{\text{t}}$) which improves the flux limit by a factor of two to three compared to the existing $n_{\text{triplet}}$ variable. Ultimately, for an energy release of $1~\text{GeV}$ for each interaction, we find that masses close to one gram and cross sections below $10^{-27}~\text{cm}^2$ (corresponding to an interaction length in ice of 10s of meters) can be probed. For larger $E_{\text{X}} \gtrsim 2~\text{GeV}$, we expect the limits to improve with the full IC86/DC detector, but a better understanding of the background using a data-driven approach is needed. For smaller $E_{\text{X}} \lesssim 0.25~\text{GeV}$, we expect better sensitivity using low-energy-threshold neutrino detectors such as DUNE and SuperK.

\vspace{0.5cm}
\subsubsection*{Acknowledgments}
We thank Sebastian Fiedlschuster, Francis Halzen, Albrecht Karle and Nicholas Orlofsky for useful discussion. The work of YB and MK is supported by the U.S.~Department of Energy under the contract DE-SC-0017647.  The work of JB is supported by the National Science Foundation under Grant No.\ 2112789. YB and JB are grateful to the Mainz Institute for Theoretical Physics (MITP) of the Cluster of Excellence PRISMA+ (Project ID 39083149), for its hospitality and its partial support during the completion of this work.


\providecommand{\href}[2]{#2}\begingroup\raggedright\endgroup

\end{document}